\documentclass[a4paper,10pt]{revtex4-2}
\usepackage{mathrsfs}
\usepackage{graphicx}
\usepackage{latexsym}
\usepackage{amsmath}
\usepackage{amssymb}
\usepackage{textcomp}
\usepackage{amsbsy}
\usepackage{graphics}
\usepackage{epstopdf}
\usepackage{color}

\allowdisplaybreaks[4]

\begin{document}

\tolerance=5000

\title{Little Rip, Pseudo Rip and bounce cosmology from generalized equation of state in the Universe with spatial curvature}

\author{A.~V.~Timoshkin,$^{1,2}$\,\thanks{alex.timosh@rambler.ru}
A.~V.~Yurov,$^{3}$\,\thanks{aiurov@kantiana.ru}
}

  \affiliation{ $^{1)}$ Institute of Scientific Research and Development, Tomsk State Pedagogical University (TSPU),  634061 Tomsk, Russia \\
$^{2)}$ Lab. for Theor. Cosmology, International Centre  of Gravity and Cosmos, Tomsk State University of Control Systems and Radio Electronics
(TUSUR),   634050 Tomsk, Russia \\
$^{3)}$ Institute of High Technology,
Baltic Federal University the name Immanuel Kant (BFU), 236041 Kaliningrad, Russia
%$^{4)}$ Department of Physics, Chandernagore College, Hooghly - 712 136.\\
%$^{(5)}$ Department of Theoretical Physics, Indian Association for the Cultivation of Science, 2A $\&$ 2B Raja S.C. Mullick Road, Kolkata - 700 %032, India
}

\tolerance=5000

\begin{abstract}

We consider the Little Rip (LR), Pseudo Rip (PR) and bounce cosmological
models in the Friedmann–Robertson–Walker (FRW) metric with nonzero
spatial curvature. We describe the evolution of the universe using a
generalized equation of state in the presence of a viscous fluid.
The conditions of the occurrence of the LR, PR and bounce were obtained from the
point of view of the parameters of the generalized equation of state
for the cosmic dark fluid, taking into account the spatial curvature.
The analytical  expressions for the spatial curvature were obtained.
Asymptotic cases of the early and late universe are considered.
A method of Darboux transformation was proposed in the case of models of an accelerating universe with viscosity.
\\

\\


\end{abstract}

%\pacs{}

\maketitle

\section{Introduction}

 As is known, a dark energy universe shows a peculiar behavior with respect to singularities in the far future,
 even in cases where it is induced by modified gravity  \cite{1} (for reviews, see \cite{2,3,4,5}).
 A general review of dark energy was given in Ref.~\cite{6}.
 Among various different possible models investigated in the literature,
 there are the models, which treat dark energy and dark matter as ideal (nonviscous) fluids with an unusual EoS.
 Very general dark fluid models with an inhomogeneous EoS were introduced in \cite{7,8,9,10}.
There exist various cosmological scenarios for the evolution of the
universe including the Big Rip \cite{11,12}, the LR \cite{13,14,15,16,17,18,19,20}, the PR \cite{21} and the
Quasi Rip (QR) \cite{22}.
The analytical representation of the cosmological models LR and PR in terms of the parameters of the generalized equation of state without taking into account the bulk viscosity of the dark fluid in the flat space was obtained in articles \cite{23,24,25,26}.
The investigation of bounce cosmological models in the flat FRW metric, taking into account the viscosity properties of a dark fluid, was carried out in the work \cite{27}). There have been proposed reconstruction works in terms of modified gravity theories including \cite{28,29,30}.

That means, the EoS will play an important role, for the LR as well
as for the PR  and the bounce phenomena. We obtain analytical expressions for the spatial curvature,
thermodynamic parameter or for the bulk viscosity in these models. In particular, we derive the
parameters for the early and late stages of the history of the universe.

The motivation of this work is that from the astronomical measurements of the luminosity of remote objects using the Planks satellite,
follows that our universe is almost flat (the spatial curvature is less than 0.03) \cite{31}.
However, studies conducted by other scientists, for example the observations of the galaxy
distribution in space, show that this conclusion is not absolute.
It is possible that the Universe will eventually have a finite spatial curvature \cite{32,33}.
The investigations, which jointly took into account the influence of bulk viscosity and spacial curvature to describe the evolution of the universe, were carried out in \cite{34,35}.

There are many articles devoted to the singularities with the final time of the formation.
The classification of the types of the singularities is proposed by Nojiri-Odintsov-Tsujikawa in \cite{1}.
However, the singularity is not the only possible ending of the evolution of the universe.
The cosmological models of the late-time universe both the LR and the PR,
the bounce cosmology, describing the cyclic universe, considered in the presented article, are non-singular.

 \section{Viscous cosmological models via an inhomogeneous fluid}

We consider a universe filled one-component ideal fluid: a dark energy, in the FRW
metric with a nonzero spatial curvature and scale factor $a$.

Let's write the energy conservation law

\begin{equation}
\dot{\rho}+3H(\rho+p)=0. \label{1}
\end{equation}

where  $H=\dot{a}/a$ is the Hubble rate and  $k^2=8\pi G$  with  $G$ being Newton's gravitational constant.
 Here $\rho,p$ are the energy density and the pressure of a dark energy
 respectively.
 A dot denotes the derivative with respect to cosmic time $t$.

We will study the homogeneous and isotropic FRW expanding universe,
\begin{equation}
ds^2 = -dt^2 + a^2(t)\left( \frac{dr^2}{1-\Pi_0 r^2}+r^2 d\Omega^2 \right), \label{2}
\end{equation}
where $d\Omega^2= d\theta^2+\sin^2\theta d\varphi^2$, $t$ is the cosmic time, $a(t)$ denotes the scale factor and has the unit of the length,
 $r$ is the special radius coordinate, and the parameter $\Pi_0$ is the spacial curvature of the three-dimensional space.

As is known, Eq.~(\ref{1}) geometrically describes different types of the universe. Taking for simplicity $\Pi_0$ to be nondimensional, for $\Pi_0 = 0$ the universe is spatially flat, for $\Pi_0 =1$ it is closed, and for $\Pi_0=-1$ it is open. The character of the universe expansion depends on the spatial curvature: the open universe will expand forever,  the flat universe will also expand forever, although at $t \rightarrow +\infty $ the expanding velocity will be constant; the closed universe will expand up to a certain instant, after which the expansion is replaced by a compression leading finally to a collapse.

The Friedmann equation for a one-component fluid in a space with nonzero spatial curvature has the form
\begin{equation}
H^2 = \frac{k^2}{3}\rho_{\rm eff}-\frac{\Pi_0}{a^2}, \label{3}
\end{equation}
where $\rho_{\rm eff}$ is the effective total energy density, $k^2=8\pi G$ with $G$ the Newtonian gravitational constant, and $H(t)= \dot{a}(t)/a(t)$ is the Hubble function.

We will use the following equation of state (EoS) for an inhomogeneous viscous fluid \cite{8}
\begin{equation}
p= \omega(\rho,t)\rho-3H\zeta(H,t), \label{4}
\end{equation}
where $\zeta(H,t)$ is the bulk viscosity, dependent on $H$ and $t$. From ordinary thermodynamics, we know that $\zeta(H,t)>0$.

We take the EoS parameter $\omega$ to have the form \cite{8}
\begin{equation}
\omega(\rho,t)=\omega_1(t)(A_0\rho^{\alpha-1} -1), \label{5}
\end{equation}
where $A_0 \neq 0$ and $\alpha \geq 1$ is a constant. A note on dimensions: as $\omega$ and $\omega_1$ are nondimensional, the dimension of $A_0$ will be complicated when $\alpha >1$. In the simplest case $\alpha =1$, however, $A_0$ will be nondimensional. Then we put $\omega(\rho,t)=\omega_0$, a constant.

Dissipative processes are described by the bulk viscosity in the form \cite{8}
\begin{equation}
\zeta(H,t)= \zeta_1(t)(3H)^n, \label{6}
\end{equation}
where $n>0$ and $\zeta_1(t)$ is an arbitrary function depending on time.

We will now consider the cosmological models of the viscous fluid.

\subsection{Little Rip model}

 Let us consider a LR model with the following form for the Hubble parameter \cite{11}:
\begin{equation}
H=H_0 \exp(\lambda t), \quad H_0>0, \lambda >0. \label{7}
\end{equation}
We will follow the development of the universe beginning at some time $t=0$ which will at first be left unspecified,
except that it refers to an initial instant in the very early universe. Its precise meaning will be dependent on which model we consider.
The symbol  $H_0$ means the Hubble parameter at this particular instant.

 In the simplest case, when $\omega(\rho,t)=\omega_0$ is a constant, and $\zeta(H,t)=\zeta_0$, with a constant $\zeta_0>0$,
the generalized equation of state takes the form:
\begin{equation}
p=\omega_0 \rho-3\zeta_0 H. \label{8}
\end{equation}
Then, using the equations (\ref{3}), (\ref{7}) and (\ref{8}), we can rewrite the energy conservation law (\ref{1}) in the view
\begin{equation}
(\omega_0+1)H^2+(\frac{2}{3}\lambda - \zeta_0 k^2)H+(\omega_0+\frac{1}{3})\frac{\Pi_0}{a^2}=0. \label{9}
\end{equation}

Solving the equation (\ref{9}) with respect to spacial curvature $\Pi_0$, we have

\begin{equation}
\Pi_0=(\omega_0+\frac{1}{3})^{-1}Ha^2[\zeta_0 k^2-\frac{2}{3}\lambda-(\omega_0+1)H]. \label{10}
\end{equation}

Note, that the Hubble function in the equation (\ref{10}) is known, it is defined in the LR model (\ref{7})
and the spatial curvature parameter accepts the values $\Pi_0=0,+1,-1$.

If $\omega_0=-1$, then
\begin{equation}
\Pi_0=(\lambda-\frac{3}{2}\zeta_0 k^2)a^{2}H. \label{11}
\end{equation}
In the case of the flat universe, when the spacial curvature $\Pi_0=0$, the thermodynamic parameter is equal
\begin{equation}
\omega_0=-1-H_0^{-1}(\frac{2}{3}\lambda - \zeta_0 k^2)\exp(-\lambda t). \label{12}
\end{equation}

Let's consider the asymptotic cases.

If $t \rightarrow 0$ (the early universe), then $\omega_0 \rightarrow  -1-H_0^{-1}(\frac{2}{3}\lambda - \zeta_0 k^2)$.
While in the late universe, when $t \rightarrow +\infty$, we obtain $\omega_0 \rightarrow -1$ asymptotically in the far future.

If the parameter of the spatial curvature $\Pi_0\neq 0$, then
\begin{equation}
\omega_0=\frac{\zeta_0 k^2-\frac{2}{3}\lambda-H_0 \exp(\lambda t)-\frac{\Pi_0}{3a_0^{2}H_0}\exp[-(\frac{2H_0}{\lambda}\exp(\lambda t)+\lambda t)]}
{H_0 \exp(\lambda t)+\frac{\Pi_0}{a_0^{2}H_0}\exp[-(\frac{2H_0}{\lambda}\exp(\lambda t)+\lambda t)]}. \label{13}
\end{equation}

The expression (\ref{13}) for the $\omega_0$ contains the correction caused by a spatial curvature $\Pi_0$.

Let's put $\lambda=H_0$, then in the asymptotic cases we obtain:

if $t \rightarrow 0$, when $\omega_0 \rightarrow \frac{\zeta_0 k^2 - \frac{5}{3}H_0 -\frac{\Pi_0}{3H_0(ea)^2}}
{H_0+\frac{\Pi_0}{H_0 a^2}}$.
On the other hand, when $t \rightarrow +\infty$, we obtain $\omega_0 \rightarrow - \frac{5}{3}+H_0^{-1}\zeta_0 k^2$ (the phantom phase)
as in the flat metric of the Friedmann universe.
The LR behavior in this case is caused by the modified condition (\ref{13}) with the parameter $\zeta_0$.

\subsection{Pseudo Rip model}

We will now investigate example in which the Hubble function approaches a constant in the far future.
That means, the universe tends asymptotically a de Sitter space.
The PR, proposed in \cite{13}, is a second variant of the theory.
This interesting possible scenario is related to the LR cosmology when the
Hubble parameter tends to infinity in the remote future \cite{14,15,16,17,18,19,20}
\begin{equation}
H(t)\rightarrow H_\infty, \quad t\rightarrow +\infty. \label{14}
\end{equation}

We will make this analysis by analogy with the LR model above.

Assume that the Hubble parameter is given as \cite{13}
\begin{equation}
H=H_0-H_1\exp(-\lambda t), \label{15}
\end{equation}
where $H_0, H_1$ and $\lambda$ are positive constants, $H_0>H_1$, and $t>0$.

Let's take the equation of state of the dark fluid again in the form (\ref{8}).
Further we will assume in (\ref{15}) that $\lambda=H_1$ and denote $\bar{H}=H_0-H_1 \exp(-H_1 t)$.
Using (\ref{3}) and (\ref{15}), we rewrite the equation of energy conservation law (\ref{1}):

\begin{equation}
\frac{\Pi_0}{a_0^2}\exp[-2(H_0 t+\exp(-H_1 t))](3H-2\bar{H})=-H[2H_0 H_1 \exp(-H_1 t)+9H(H+\zeta_0 k^2)], \label{16}
\end{equation}

Here the Hubble function is defined in (\ref{15}).

In the case of the flat universe $\Pi_0=0$  the equation (\ref{16}) is simplified, we get
\begin{equation}
2H_0 H_1 \exp(-H_1 t)+9H(H+\zeta_0 k^2)=0, \label{17}
\end{equation}

Let's express from here the thermodynamic parameter
\begin{equation}
\omega_0=-1-H^{-1}[\frac{2H_0 H_1}{9H}\exp(-H_1 t)+\zeta_0 k^2] \label{18}
\end{equation}

It take place the phantom phase ($\omega_0<-1$) of the evolution of the universe, but the singularity is not formed.

Let's consider the asymptotic cases.

In the early universe, when $t \rightarrow 0$, $\omega_0 \rightarrow  -1-H_0^{-1}[\frac{2H_0 H_1}{9(H_1-H_0)}+\zeta_0 k^2]$.
In the far future $t \rightarrow +\infty$, we get $\omega_0 \rightarrow  -1-\frac{\zeta_0 k^2}{H_1}$.
We see that the PR is determined by the parameter $\zeta_0$.

Let's consider the case, when the parameter of the spatial curvature $\Pi_0\neq 0$.

Solving Eq. (\ref{16})  with respect to $\zeta_0$, we have

\begin{equation}
\zeta_0=(9k^2H)^{-1}[\frac{\Pi_0 (2H-3\bar{H})}{a_0^2 H}\exp[-2(H_0 t+\exp(-H_1 t)]-2H_0 H_1 \exp(-H_1 t)]-\frac{(\omega_0+1)H}{k^2}, \label{19}
\end{equation}

The first term in the expression (\ref{19}) takes into account the contribution of the spatial curvature.

Thus, the PR behavior in this model is determined by the modified condition (\ref{19}) with the thermodynamic parameter $\omega_0$.

\subsection{Bounce cosmology}

In this model the universe goes from an era of accelerated collapse to the expanding era without displaying
a singularity. This is essentially a cyclic universe model. After the bounce the universe soon enters a matter dominated
expansion phase \cite{36,37,38}. In the following we will study cosmological models of basically this type.

\emph{First example}

Let us consider a bounce cosmological model where the scale factor $a(t)$ has an exponential form \cite{37}:

\begin{equation}
a(t)=a_0 \exp[\alpha(t-t_0)^{2n}], \quad n \in N, \label{20}
\end{equation}

where $\alpha$ is a positive (dimensional) constant and $t_0$ is the instant when bouncing occurs.

The Hubble function becomes
\begin{equation}
H(t)=2n\alpha(t-t_0)^{2n-1}. \label{21}
\end{equation}

Taking into account Eq. (\ref{1}), (\ref{3}), (\ref{8}), (\ref{20}) and (\ref{21}) we can write the energy conservation law in the form

\begin{equation}
\Pi_0=\frac{2n \alpha a_0^2 (t-t_0)^{2n-1} \exp2\alpha(t-t_0)^{2n}}{\omega_0+1}[\zeta_0 k^2-\frac{2}{3}(2n-1)(t-t_0)^{-1}-2n \alpha (\omega_0+1) (t-t_0)^{2n-1}], \label{22}
\end{equation}

where $\omega_0\neq -1$.

In the simplest case $n=1$ we obtain

\begin{equation}
\Pi_0=\frac{2 \alpha a_0^2 (t-t_0) \exp2\alpha(t-t_0)^2}{\omega_0+1}[\zeta_0 k^2-\frac{2}{3}(t-t_0)^{-1}-2 \alpha (\omega_0+1) (t-t_0)], \label{23}
\end{equation}

If $\Pi_0=0$ one can formulate bounce cosmology theory in terms of the bulk viscosity concept.

\begin{equation}
\omega_0=-1-\frac{1}{2\alpha(t-t_0)}[\zeta_0 k^2-\frac{2}{3(t-t_0)}]. \label{24}
\end{equation}

This is the thermodynamic parameter, corresponding to the exponential model.

If $\Pi_0\neq 0$, then from the equation (\ref{23}) follows the representation of an exponential model in terms of the bulk viscosity

\begin{equation}
\omega_0=-1+ \frac{\alpha \zeta_0 k^2}{2\alpha^{2}(t-t_0)+\frac{\Pi_0}{2a_0^2(t-t_0)}\exp[-2\alpha(t-t_0)^2]}. \label{25}
\end{equation}

The expression (\ref{25}) contains in the denominator the correction caused by spatial curvature $\Pi_0$.

Thus, we have modified condition for the appearance of the bounce cosmology when the Hubble function has an exponential form.

\emph{Second example}

Power-law model. Now we will analyze the bounce cosmological scenario when the scale factor has the following form \cite{37}:

\begin{equation}
a(t)=a_0+\alpha(t-t_0)^{2n}, \label{26}
\end{equation}

where $\alpha$ is dimensional constant, $n \in N$ and $t=t_0$ is a fixed bounce time.

The Hubble function becomes

\begin{equation}
H(t)=\frac{2n\alpha(t-t_0)^{2n-1}}{a_0+\alpha(t-t_0)^{2n}}. \label{27}
\end{equation}

Let's consider the case $n=1$.

Taking into account (3), (8), (26) and (27) from the  energy conservation law should be

\begin{equation}
\Pi_0=\frac{2\alpha}{\omega_0+1}[\zeta_0 k^2 (t-t_0)[a_0+\alpha(t-t_0)^{2}]-\frac{2}{3}[a_0-\alpha(t-t_0)^{2}]-2\alpha(\omega_0+1)(t-t_0)^{2}]. \label{28}
\end{equation}

If the universe is spatially flat, that corresponds to $\Pi_0=0$, then from the equation (\ref{28}) we obtain

\begin{equation}
\omega_0=-1+\frac{1}{2}[\frac{a_0\zeta_0k^2}{\alpha(t-t_0)}+\zeta_0k^2(t-t_0)-\frac{a_0}{3\alpha(t-t_0)^2}+\frac{1}{3}]. \label{29}
\end{equation}

This is the thermodynamic parameter, corresponding to the power-law model.

On the other hand, when the spatial curvature is different from zero $\Pi_0\neq0$ (the universe is of an open or a closed type),
when $t\rightarrow t_0$, near the bounce time $t_0$, a similar idea looks like this

\begin{equation}
\omega_0\rightarrow -1-\frac{4\alpha}{\Pi_0}. \label{30}
\end{equation}

Thus, a generalization of the theory with the LR, PR and bounce cosmology was obtained,
taking into account the bulk viscosity of the dark fluid in the space of FRW with nonzero spatial curvature.
We formulated the modified conditions for the appearance of the presented cosmological models due to the nonzero spatial curvature of the space-time.

\section{Darboux transformation}
In this section, we will show that the Darboux transform is an efficient tool for constructing exact solutions in the case of viscous cosmological models. First of all, we write the equation for the second derivative of the scale factor. To do this, we differentiate (3), use (1), and after simple transformations one get
\begin{equation}
\ddot{a}=ua+\frac{3k^2}{2}\zeta\dot{a},
\label{7.1}
\end{equation}
where  we introduced the compact notation
\begin{equation}
u=u(t)\equiv -\frac{k^2}{6}\rho(1+3\omega).
\label{9.1}
\end{equation}
To define the Darboux transformation (DT) let's introduce an auxiliary function $f(t)$, which will be referred to as the {\em support function}. The support function is the solution of the auxiliary linear equation
\begin{equation}
\ddot{f}=(u+c)f+\frac{3k^2}{2}\zeta\dot{f},
\label{8.1}
\end{equation}
where $c$=const will be referred as the spectral parameter. In this sense, the scale factor $a(t)$ is the solution of the equation (\ref{7.1}) with zero value of the spectral parameter.

Let us define the function $\sigma={\dot f}/f$  that will satisfy the Ricatti-type equation
\begin{equation}
{\dot \sigma}=u+c+\frac{3}{2}k^2\zeta\sigma-\sigma^2.
\label{sig}
\end{equation}
Now, by direct calculation, one can verify the validity of the following theorem:

{\bf Theorem.} Let $\theta=\theta(t)$ is arbitrary differentiable function, $a$ and $f$ are solutions of the (\ref{7.1}) and (\ref{8.1}). Then the equation (\ref{7.1}) is covariant with respect to the following transformation:
\begin{equation}
\begin{cases}
a\to a^{(1)}={\rm e}^{\theta}\left({\dot a}-\sigma a\right),\\
u\to u^{(1)}=u+\frac{3}{2}k^2\left({\dot\zeta}-\zeta{\dot\theta}\right)+{\ddot\theta}-{\dot\theta}^2-2{\dot\sigma},\\
\zeta\to \zeta^{(1)}=\zeta+\frac{4}{3k^2}{\dot\theta}
\end{cases} \label{DT}
\end{equation}
In other words, the function  $a^{(1)}$  is a solution of  the equation
\begin{equation}
{\ddot{a}}^{(1)}=u^{(1)}a^{(1)}+\frac{3k^2}{2}\zeta^{(1)}{\dot{a}}^{(1)}.
\label{7.11}
\end{equation}
The validity of the theorem is verified by direct calculation. The transformation (\ref{DT})  is called Darboux transformation (DT).

{\em Comment}. A similar theorem can be proved for an arbitrary power of the scale factor $a^m$, but only for the flat case $\Pi_0=0$. For non-zero curvature, the theorem only works for the first power of the scale factor. Since in this paper we are investigating precisely the situation with $\Pi_0\ne 0$, we use only this power of the scale factor in our theorem.

Using (\ref{DT}) one can find exact expressons for the $\rho^{(1)}$, $\omega^{(1)}$ and so on. It is important to note, that the number of free functions in the original and transformed expressions is sufficient to satisfy all Einstein's equations, not just (\ref{7.11}).

Unfortunatel, the general expressions for the transformed quantities are extremely cumbersome, so we will restrict ourselves to the formulas for the "dressed" density $\rho^{(1)}$  and the Hubble parameter $H^{(1)}$:
\begin{equation}
H^{(1)}=-Q-\frac{c}{q}+\frac{3k^2\zeta}{2},
\label{H11}
\end{equation}
and
\begin{equation}
\rho^{(1)}=\frac{3}{k^2q^2}\left(c^2+\frac{\Pi_0^{(1)}{\rm e}^{-2\theta}}{a^2}\right)+\frac{3Q}{k^2}\left(Q+\frac{2c}{q}\right)+9\zeta\left(\frac{3k^2\zeta}{4}-Q-\frac{c}{q}\right),
\label{r1}
\end{equation}
where
\begin{equation}
q=H-\sigma,\qquad Q=\sigma-{\dot\theta}.
\label{qQ}
\end{equation}

In the case of general position, DT is used as follows: an equation   (\ref{7.1})  with a known exact solution is considered, after which the formulas (\ref{DT})  are applied and more complex and nontrivial solutions are obtained. To illustrate the effectiveness of the Darboux transformation, consider the simple case of form-invariant potentials, i.e. potentials trivially transforming under this transformation.

Let
$$
\theta=0,\qquad \zeta=\zeta_0={\rm const},
$$
and
\begin{equation}
f(t)=f_0\exp\left(\frac{\xi t^2}{4}\right),\qquad \xi={\rm const}.
\label{f0}
\end{equation}
Then, the transformation (\ref{DT}) results in
\begin{equation}
a^{(1)}={\dot a}-\frac{\xi t}{2}a,\qquad \zeta^{(1)}=\zeta_0,\qquad u^{(1)}=u-\xi.
\label{DT0}
\end{equation}
The exact form of the support function  (\ref{f0}) was chosen specifically so that the transformation of the potential $u\to u^{(1)}$ would be reduced to adding a constant $\xi$. Similarly, in the case of the Schrodinger equation, the harmonic oscillator potential behaves, which is the most famous example of a shape-invariant potential.

Substituting (\ref{f0}) in the (\ref{8.1}) one get the initial potential $u(t)$:
\begin{equation}
u=\frac{\xi^2}{4}\left(t-\frac{3k^2\zeta_0}{2\xi}\right)^2+\frac{\xi}{2}-\left(\frac{3k^2\zeta_0}{4}\right)^2-c.
\label{u0}
\end{equation}
Using (\ref{u0}) one can solve the equation (\ref{7.1}) to find the scale fuctor $a(t)$ (which will be denote as $a_0(t)$) and the spectral parameter $c$ (which will be denote as $c_0$).  One get
\begin{equation}
\begin{cases}
a_0(t)=\exp\left[\frac{\kappa}{4\xi}\left(\xi t-\frac{3\zeta_0 k^2}{2}\right)^2+\frac{3\zeta_0 k^2 t}{4}\right],\\
c_0=\frac{1}{2}(1-\kappa)\xi,\qquad \kappa=\pm 1.
\end{cases}
\label{a0}
\end{equation}
and the Hubble parameter
\begin{equation}
H_0=\frac{\kappa\xi t}{2}-\frac{3\zeta_0 k^2(\kappa-1)}{4}.
\label{H0}
\end{equation}
Thus we have LR if $\kappa=+1$ with  density
\begin{equation}
\rho_0=\frac{3}{k^2}\left[\frac{\xi^2t^2}{4}+\Pi_0\exp\left(-\frac{9k^4\zeta_0^2+4\xi^2t^2}{8\xi}\right)\right].
\label{rr0}
\end{equation}
It is interesting ti note that after the DT (\ref{DT}) we have new scale factor
\begin{equation}
a^{(1)}(t)=\frac{a_0(t)}{4}(2\xi t-3\zeta_0k^2)(\kappa-1),
\label{a01}
\end{equation}
so if $\kappa=+1$ then $a^{(1)}=0$.

Of course, (\ref{a0})  is only a particular solution of the equation (\ref{7.1}). In general, the solution can be expressed in terms of Hermite polynomials. We give the following two simple cases
\begin{equation}
\begin{cases}
a_1(t)=\left(\xi t-\frac{3k^2\zeta_0}{2}\right)a_0(t),\\
c_1=\frac{1}{2}(1-3\kappa)\xi,\qquad \kappa=\pm 1,
\end{cases}
\label{a1}
\end{equation}
and
\begin{equation}
\begin{cases}
a_2(t)=\left[\left(\xi t-\frac{3k^2\zeta_0}{2}\right)^2+\frac{\xi}{\kappa}\right]a_0(t),\\
c_2=\frac{1}{2}(1-5\kappa)\xi,\qquad \kappa=\pm 1,
\end{cases}
\label{a2}
\end{equation}
with $a_0$ from the equation (\ref{a0}). We will not write down the general expression for $a_n(t)$, but will restrict ourselves to the general form of the spectral parameter
\begin{equation}
c_n=\frac{\xi}{2}\left(1-(2n+1)\kappa\right),\qquad n=0,\,1,\, 2....,
\label{cn}
\end{equation}
we only note that all these expressions for the scale factor are related to each other by the Darboux transformation
\begin{equation}
a^{(1)}_{n+1}(t)=a_n(t),
\label{n1n}
\end{equation}
which reflects the shape-invariance property (In the case of a harmonic oscillator, these transformations act as annihilation operators).

The situation becomes more interesting if we use the function $\theta$. We start with expressions (\ref{u0}), (\ref{a0})  but with
\begin{equation}
\theta=\exp\left(\lambda t \right).
\label{th}
\end{equation}
DT (\ref{DT}) results in

\begin{equation}
\begin{cases}
a^{(1)}_0(t)= \exp\left({\rm e}^{\lambda t}\right)(2\xi t-  3k^2\zeta_0)a_0(t),\\

\zeta^{(1)}=\zeta_0+\frac{4\lambda}{3 k^2}{\rm e}^{\lambda t}.
\end{cases}
\label{ath}
\end{equation}

As follows from (\ref{ath}), the resulting solution describes a Little Rip cosmology  with an exponentially growing Hubble parameter:

\begin{equation}
H^{(1)}=\lambda {\rm e}^{\lambda t}+\frac{2\xi}{2\xi t-3 k^2\zeta_0}+H_0,
\label{H1}
\end{equation}
where $H_0$ is defined by the equation (\ref{H0}).

As we have already said, the presence of free functions $\omega_1(t)$ in (\ref{5}) and $\zeta_1(t)$ in (\ref{6}) (we will call them gauge functions)  allows us to show that the quantities transformed by applying the Darboux transformation (\ref{DT}) are indeed solutions of the entire system of Einstein-Friedmann cosmological equations. Let's show this with our example. First at all, take into account that  the density $\rho^{(1)}$, Hubble root $H^{(1)}$ (\ref{H1}) and scale factor $a{(1)}_0$ (\ref{ath})  satisfy the equation (\ref{3})  with, generally speaking, the new spatial curvature $\Pi^{(1)}_0$:
\begin{equation}
\left(H^{(1)}\right)^2 = \frac{k^2}{3}\rho^{(1)}-\Pi^{(1)}_0 \left(a^{(1)}\right)^{-2}. \label{3+}
\end{equation}
Substituting (\ref{ath}) and (\ref{H1}) into the (\ref{3+}) one can  express the function $\rho^{(1)}$. In the next step, we calculate the function $u^{(1)}$ from the (\ref{DT}) and substitute $u^{(1)}$ and $\rho^{(1)}$ into the dressed (\ref{9.1}) to calculate $\omega^{(1)}$:
\begin{equation}
\omega^{(1)}=-\frac{1}{3}-\frac{2u^{(1)}}{k^2 \rho^{(1)}}.
\label{om1}
\end{equation}
In order for our calculations to be consistent, we impose the connection condition (\ref{5}) on $\rho^{(1)}$  and $\omega^{(1)}$, which allows us to calculate the transformed gauge function $\omega^{(1)}_1(t)$:
\begin{equation}
\omega^{(1)}_1(t)=\frac{\omega^{(1)}}{A_0 \left(\rho^{(1)}\right)^{\alpha-1}-1},
\label{om11}
\end{equation}
with $\omega^{(1)}$ from the (\ref{om1}).

Similarly, one should fix the gauge arbitrariness determined by the function $\zeta_1(t)$. Using (\ref{ath}), (\ref{H1}) and (\ref{6}) one get for the case $\kappa=+1$ the gauge condition:
\begin{equation}
\zeta^{(1)}_1(t)=\left(\zeta_0+\frac{4\lambda}{3 k^2}{\rm e}^{\lambda t}\right)\left(\lambda {\rm e}^{\lambda t}+\frac{2\xi}{2\xi t-3 k^2\zeta_0}+\frac{\xi t}{2}\right)^{-n}.
\label{z11}
\end{equation}

Thus, by DT, a method of generating exact decisions was proposed in the case of models of an accelerating universe with viscosity.

\section{Conclusion}

In this investigation we have considered the cosmological models for the LR, PR and bounce universes in the FRW space-time with a nonzero spatial curvature. We have shown, that these cosmological scenarios are determined by the parameters $\omega_0$, $\zeta_0$ and $\Pi_0$.
We have modified the representation of the LR, PR and bounce theories in terms of the parameters of the generalized EoS, taking into account the parameter of the spatial curvature $\Pi_0$.

The main emphasis in this study is that the cosmic space has nonzero spatial curvature.
It may also be mentioned that the inclusion in the Friedmann equation of a term with a nonzero spatial curvature allows you to get a more detailed description of the evolution of the universe.

The cosmological analysis for the cases of an open and a closed late-time universe shows, that the thermodynamic parameter $\omega_0$ of a generalized EoS takes the value of less than a minus one (the phantom phase), similarly to a spatial flat universe. However, despite this fact, in the models of the late-time universe with the LR and PR there is an accelerated expansion process without the occurrence of the singularities. Physically this means, that when we describe the gravity on the cosmological scale, the value of the spatial curvature parameter weakly affects on the nature of the evolution of the universe.

It has been proved the theorem about the DT, allows to find accurate decisions of the differential equation for a scale factor. As the application of this theorem, in the first approximation the expressions for the Hubble function, the energy density and the thermodynamic parameter in the LR model with nonzero spatial curvature were obtained.

 It is interesting that within this scenario one can consider the unified bounce with dark energy epoch (see examples in \cite{39}).
The description in a unifying way the early and the late-time universe, when the generalized EoS contains a bulk viscosity was considered in \cite{40}).
\\

\textbf{Acknowledgment}
\\

The article was supported by the Ministry of Science and Higher Education of the Russian Federation (agreement no. 075-02-2023-934)

\end{document}